\documentclass[twocolumn,tighten,astrosymb,trackchanges]{aastex631}
\newcommand{\LJMU}{\affiliation{Astrophysics Research Institute, Liverpool John Moores University, 146 Brownlow Hill, Liverpool L3 5RF, UK}}

\newcommand{\birmingham}{\affiliation{School of Physics \& Astronomy and Institute of Gravitational Wave Astronomy, University of Birmingham, UK}}

\newcommand{\DIRAC}{\affiliation{DIRAC Institute, Department of Astronomy, University of Washington, 3910 15th Avenue NE, Seattle, WA 98195, USA}}

\newcommand{\HUBerlin}{\affiliation{Institut f\"ur Physik, Humboldt-Universit\"at zu Berlin, Newtonstr. 15, 12489 Berlin, Germany}}

\newcommand{\chapelhill}{\affiliation{Department of Physics and Astronomy, University of North Carolina at Chapel Hill, Chapel Hill, NC 27599-3255, USA}}

\newcommand{\OKC}{\affiliation{Department of Physics, Oskar Klein Centre, Stockholm University, SE-106 91, Stockholm, Sweden}}
\newcommand{\OKCAstro}{\affiliation{Department of Astronomy, Oskar Klein Center, Stockholm University, SE-106 91 Stockholm, Sweden}}
\newcommand{\Caltech}{\affiliation{Cahill Center for Astronomy and Astrophysics, California Institute of Technology, Mail Code 249-17, Pasadena, CA 91125, USA}}
\newcommand{\CaltechPhys}{\affiliation{Division of Physics, Mathematics and Astronomy, California Institute of Technology, Pasadena, CA 91125, USA}}
\newcommand{\CaltechDDD}{\affiliation{Center for Data Driven Discovery, California Institute of Technology, Pasadena, CA 91125, USA}}

\newcommand{\UCB}{\affiliation{Department of Astronomy, University of California, Berkeley, CA 94720-3411, USA}}

\newcommand{\IPAC}{\affiliation{IPAC, California Institute of Technology, 1200 E. California Blvd, Pasadena, CA 91125, USA}}

\newcommand{\CIERA}{\affiliation{Center for Interdisciplinary Exploration and Research in Astrophysics (CIERA), 1800 Sherman Ave., Evanston, IL 60201, USA}}
\newcommand{\Northwestern}{\affiliation{Department of Physics and Astronomy, Northwestern University, 2145 Sheridan Rd, Evanston, IL 60208, USA}}
\newcommand{\SkAI}{\affiliation{NSF-Simons AI Institute for the Sky (SkAI), 172 E. Chestnut St., Chicago, IL 60611, USA}}

\newcommand{\CaltechOO}{\affiliation{Caltech Optical Observatories, California Institute of Technology, Pasadena, CA 91125, USA}}

\newcommand{\mcwilliams}{     McWilliams Center for Cosmology and Astrophysics,     Department of Physics,     Carnegie Mellon University,     5000 Forbes Avenue, Pittsburgh, PA 15213, USA }
\newcommand{\LMU}{\affiliation{University Observatory, Faculty of Physics, Ludwig-Maximilians-Universität, Scheinerstr. 1, 81679 Munich, Germany}}
\newcommand{\ORIGINS}{\affiliation{Excellence Cluster ORIGINS, Boltzmannstr. 2, 85748 Garching, Germany}}
\newcommand{\UCLA}{\affiliation{Department of Physics and Astronomy, UCLA PAB 430 Portola Plaza Los Angeles, CA 90095-1547}}
\usepackage{CJK}
\usepackage{multirow}

\let\tablenum\relax
\usepackage{siunitx}
\usepackage{appendix}
\usepackage{amsmath,amsthm,amsfonts,amssymb,amscd}
\usepackage{booktabs}
\usepackage{tablefootnote}
\usepackage{apjfonts}
\usepackage{lipsum}
\usepackage{changepage}

\let\ts=\thinspace
\newcommand{\one}{\ts {\sc i}}
\newcommand{\two}{\ts {\sc ii}}
\newcommand{\three}{\ts {\sc iii}}
\newcommand{\four}{\ts {\sc iv}}
\newcommand{\five}{\ts {\sc v}}
\newcommand{\six}{\ts {\sc vi}}
\newcommand{\seven}{\ts {\sc vii}}
\renewcommand\ion[2]{#1\,\,{\sc{\romannumeral #2}}}

\definecolor{maroon}{rgb}{0.760,0.118,0.337}

\def\cm{\mbox{\,cm}}
\def\cm3{\mbox{\,cm$^{-3}$}}

\shorttitle{Discovery of SN 2025wny}
\shortauthors{Johansson et al.}

\begin{document}

\title{Discovery of SN 2025wny: a Strongly Gravitationally Lensed Superluminous Supernova at $z=2.01$}

\correspondingauthor{Joel Johansson}
\email{joeljo@fysik.su.se}

\author[0000-0001-5975-290X]{Joel~Johansson}
\OKC

\author[0000-0001-8472-1996]{Daniel~A.~Perley}
\LJMU

\author[0000-0002-4163-4996]{Ariel~Goobar}
\OKC

\author[0000-0003-0733-2916]{Jacob~L.~Wise}
\LJMU

\author[0000-0003-3658-6026]{Yu-Jing~Qin}
\Caltech

\author[0009-0006-0726-1328]{Zoë~McGrath}
\LJMU

\author[0000-0001-6797-1889]{Steve~Schulze}
\CIERA

\author[0000-0003-2456-9317]{Cameron~Lemon}
\OKC

\author[0000-0002-3884-5637]{Anjasha~Gangopadhyay}
\OKCAstro

\author[0009-0004-1062-8886]{Konstantinos~Tsalapatas}
\OKCAstro

\author[0000-0002-8977-1498]{Igor~Andreoni}
\chapelhill

\author[0000-0001-8018-5348]{Eric~C.~Bellm}
\DIRAC

\author[0000-0002-9870-5695]{Joshua~S.~Bloom}
\UCB

\author[0000-0002-5884-7867]{Richard~Dekany}
\CaltechOO

\author[0000-0002-2376-6979]{Suhail~Dhawan}
\birmingham

\author[0000-0002-4223-103X]{Christoffer~Fremling}
\CaltechPhys
\CaltechOO

\author[0000-0002-3168-0139]{Matthew~J.~Graham}
\CaltechPhys

\author[0000-0001-5668-3507]{Steven~L.~Groom}
\IPAC

\author[0000-0003-3270-7644]{Daniel~Gruen}
\LMU
\ORIGINS

\author[0000-0002-9364-5419]{Xander~J.~Hall}
\affiliation{\mcwilliams}

\author[0000-0002-5619-4938]{Mansi~Kasliwal}
\Caltech

\author[0000-0003-2451-5482]{Russ~R.~Laher}
\IPAC

\author[0000-0001-9454-4639]{Ragnhild~Lunnan}
\OKCAstro

\author[0000-0003-2242-0244]{Ashish~A.~Mahabal}
\CaltechPhys
\CaltechDDD

\author[0000-0001-9515-478X]{Adam~A.~Miller}
\CIERA
\Northwestern
\SkAI

\author[0000-0002-8380-6143]{Edvard~Mörtsell}
\OKC

\author[0000-0001-8342-6274]{Jakob~Nordin}
\HUBerlin

\author[0009-0009-6243-8300]{Jacob~Osman~Hjortlund}
\OKC

\author[0000-0003-0427-8387]{R.~Michael~Rich}
\UCLA

\author[0000-0002-0387-370X]{Reed~L.~Riddle}
\CaltechOO

\author[0000-0003-2091-622X]{Avinash~Singh}
\OKCAstro

\author[0000-0003-1546-6615]{Jesper~Sollerman}
\OKCAstro

\author[0000-0001-6343-3362]{Alice~Townsend}
\HUBerlin

\author[0000-0003-1710-9339]{Lin~Yan}
\CaltechOO

\begin{abstract}
We present the discovery of SN 2025wny (ZTF25abnjznp/GOTO25gtq) and spectroscopic classification of this event as the first gravitationally lensed Type I superluminous supernovae (SLSN-I). 
Deep ground-based follow-up observations resolves four images of the supernova with $\sim 1.7$" angular separation from the main lens galaxy, each coincident with the lensed images of a background galaxy seen in archival imaging of the field. 
Spectroscopy of the brightest point image shows narrow features matching absorption lines at a redshift of $z=2.011$ and broad features matching those seen in superluminous SNe with Far-UV coverage.
We infer a magnification factor of $\mu \sim$20--50 for the brightest image in the system, based on photometric and spectroscopic comparisons to other SLSNe-I. SN\,2025wny demonstrates that gravitationally-lensed SNe are in reach of ground-based facilities out to redshifts far higher than what has been previously assumed, and provide a unique window into studying distant supernovae, internal properties of dwarf galaxies, as well as for time-delay cosmography.
\end{abstract}


\vspace{2.0cm}
\section{Introduction \label{sec:intro}}
Gravitationally lensed supernovae (glSNe) provide unique laboratories for both astrophysics and cosmology, offering direct probes of galaxy-scale mass distributions and independent measurements of the Hubble constant, H$_0$, through time-delay cosmography \citep{refsdal_1964b}. Over the past decade, the number of known lensed SNe has grown from single prototypes to a small but diverse sample spanning multiple SN types and redshifts. 
The first gravitationally lensed SN to be detected, PS1-10afx, was identified serendipitously and first suggested to be an exotic type of superluminous supernova \citep{2013ApJ...767..162C}, before \citet{quimby_extraordinary_2013} demonstrated it was in fact a highly magnified normal Type Ia SN.  The event had faded before multiple images could be resolved.
The first multiply imaged SN discovered, SN Refsdal, 
revealed multiple images of the same explosion due to deflection by a galaxy cluster lens \citep{kelly_2015, Kelly_2016}. The first resolved multiply-imaged Type Ia SN, iPTF16geu, was found in a galaxy-scale lens system and demonstrated the power of wide-field time-domain surveys to uncover such rare events \citep{goobar_2017}. Subsequent discoveries have expanded this class, including the strongly magnified SN Zwicky \citep{goobar_2023} and additional examples uncovered through systematic searches with HST and JWST \citep{rodney_2021, pierel_2024, frye_2024}. 
The challenges of ground-based strongly lensed SN searches are discussed in \citet{2025RSPTA.38340123G} and many of the science potentials are highlighted in the review by \citet{suyu_2024a}.

Because of line blanketing and photospheric cooling, most SNe emit little UV luminosity after the first few days of explosion \citep{Brown+2009} and are thus difficult to detect and study at distances substantially beyond $z>1$ in the optical band, even if gravitationally lensed. As a result, higher-$z$ lensing systems have remained largely the domain of space telescopes with sensitive NIR capabilities (the Hubble Space Telescope and James Webb Space Telescope), which have found many lensed SNe---typically behind known lensing clusters---but rarely classify them (e.g., \citealt{2025ApJ...979..250D, 2025ApJ...990...31D}).

In this letter we report the discovery of SN\,2025wny at $z=2.011$, a Type I superluminous supernova (SLSN) that is strongly lensed and multiply imaged by a foreground galaxy at $z=0.357$.  It is the first known lensed SLSN, and the first SN of any type for which multiple images can be separately resolved in seeing-limited images from the ground.  This event adds to the growing sample of gravitationally lensed transients and hugely expands the redshift horizon for ground-based lensing studies. It will provide new opportunities to constrain the lens potential, measure time delays for cosmography, and explore the properties of high-redshift SNe magnified by gravitational lensing in tandem with their star-forming dwarf host galaxies.

\section{Observations \label{sec:obs}}

\subsection{Discovery of SN\,2025wny}

SN\,2025wny was first identified by the Zwicky Transient Facility (ZTF; \citealt{Bellm+2019,Graham+2019,Dekany+2020,Masci+2019,Patterson+2019,Mahabal+2019,Duev+2019}) as a candidate transient on UT 2025-08-29 and given the identifier ZTF25abnjznp; forced photometry on recent images recovered prior detections dating back to 2025-08-27. It was reported to the TNS by the Gravitational-wave Optical Transient Observer (GOTO) on UT 2025-09-01 as GOTO25gtq \citep{2025TNSTR3492....1O}.

The event was identified as a candidate of interest within a day of discovery during our daily screening of new transients cross-matched against public spectroscopic and photometric redshift catalogs\footnote{{\texttt{Fritz/Skyportal} was used to facilitate source crossmatching, data sharing, and exploratory analysis \citep{vanderWalt2019,Coughlin+2023}.
}}. In this case, the transient was nearby in projection to a massive luminous red galaxy (LRG) with an archival spectrum from the Dark Energy Spectroscopic Instrument (DESI), placing the redshift at $z=0.3754$ \citep{desi2025}. The brightness of the object was inconsistent with a Type Ia SN at the redshift of the galaxy, indicating that a background source with substantial lensing magnification was viable. 

In addition to the DESI redshift, the transient was also flagged as lying $1\farcs0$ from the known lens candidate PS1J0716+3821 \citep{2020A&A...644A.163C} by our daily cross matching to the Strong Lensing Database (SLED\footnote{A cone search of 50$\arcsec$ of the SLED database: https://sled.amnh.org/}) of confirmed and candidate gravitational lenses, providing further support that SN 20205wny could be lensed.

\begin{figure*}
    \centering
    \includegraphics[width=\textwidth]{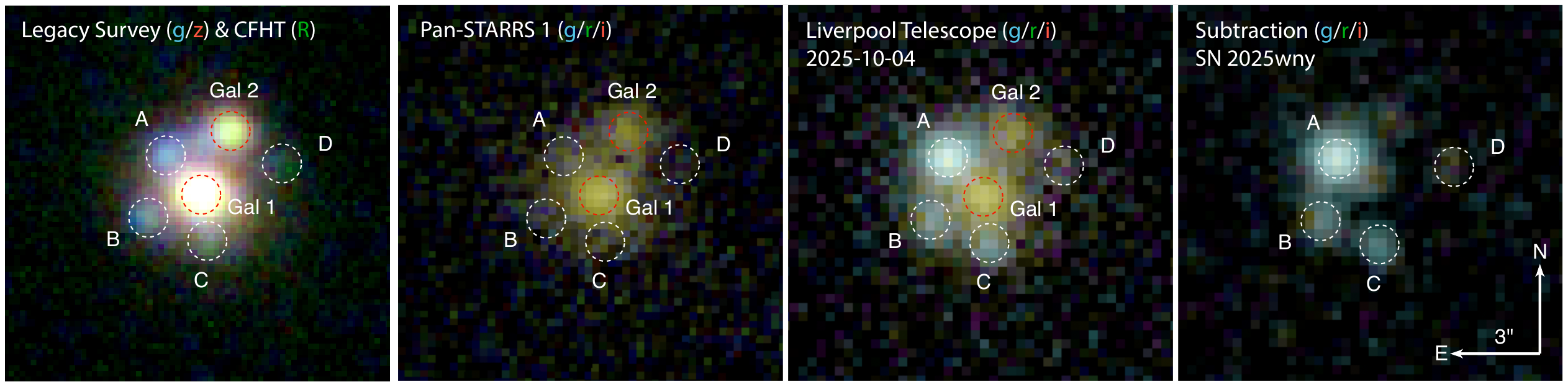}
    \caption{RGB composite images of the lens system from Legacy Survey ($g$ and $z$-bands) and CFHT $R$-band imaging (left panel) prior to the SN explosion. The following panels show the Pan-STARRS images used for image subtraction, LT $gri$ images from 2025 October 4, and the right panel the four transient images of SN\,2025wny.
    }
    \label{fig:image}
\end{figure*}

Legacy Survey imaging \citep{2019AJ....157..168D} of the field shows a second red galaxy nearby, likely at the same redshift.  Archival Canada France Hawaii Telescope (CFHT) imaging \citep{Gwyn2008PASP..120..212G} shows four images of a blue background object in a cross pattern around the LRGs and possible arc-like structures close to the brightest image.  (A colorized image combining Legacy Survey and CFHT imaging is shown in the leftmost panel of Figure \ref{fig:image}, with the four candidate images labeled A--D.)

\subsection{Imaging \label{sec:obs_imaging}}

ZTF carried out regular survey observations of the field containing SN\,2025wny starting 11 days prior to the time of discovery, which were processed according to the ZTF standard pipeline \citep{Masci+2019}. The public photometry is presented here and is available via all public ZTF brokers.

Imaging observations with the IO:O camera on the Liverpool Telescope (LT) began on 2025-09-02 UT, as soon as the potential lensing nature of the system was identified. These initial exposures, and others on subsequent nights, showed a clear detection of a point source at the location of Image A, but due to the relatively short integration time no other high-S/N sources are apparent in the image or a difference subtraction of reference imaging in the same bands from Pan-STARRS 1 \citep{chambers_pan-starrs1_2016}. However, improved conditions on 2025-09-20 and on several subsequent nights revealed multiple clear detections of three residual sources (and a marginal detection of a fourth) in the subtraction image in an Einstein Cross pattern, with locations consistent with the four candidate galaxy-lens images in the CFHT images, a result we reported to the community as an AstroNote \citep{wise2025}.  RGB false-colour images showing pre-explosion Pan-STARRS imaging, post-explosion LT imaging, and the resulting subtraction image are shown in Figure \ref{fig:image} (right three panels).
From our observations on 2025-10-04 UT, we measure the brightest SN image (A) to have magnitudes of $g = 20.42 \pm 0.04$, $r = 19.60 \pm 0.03$, $i = 19.54 \pm 0.03$.

Additional imaging observations from the Liverpool Telescope, the Palomar 60-inch Telescope, the Fraunhofer Telescope at Wendelstein Observatory, and other facilities will be presented in future work. 

\begin{figure*}
    \centering
    \includegraphics[width=\textwidth]{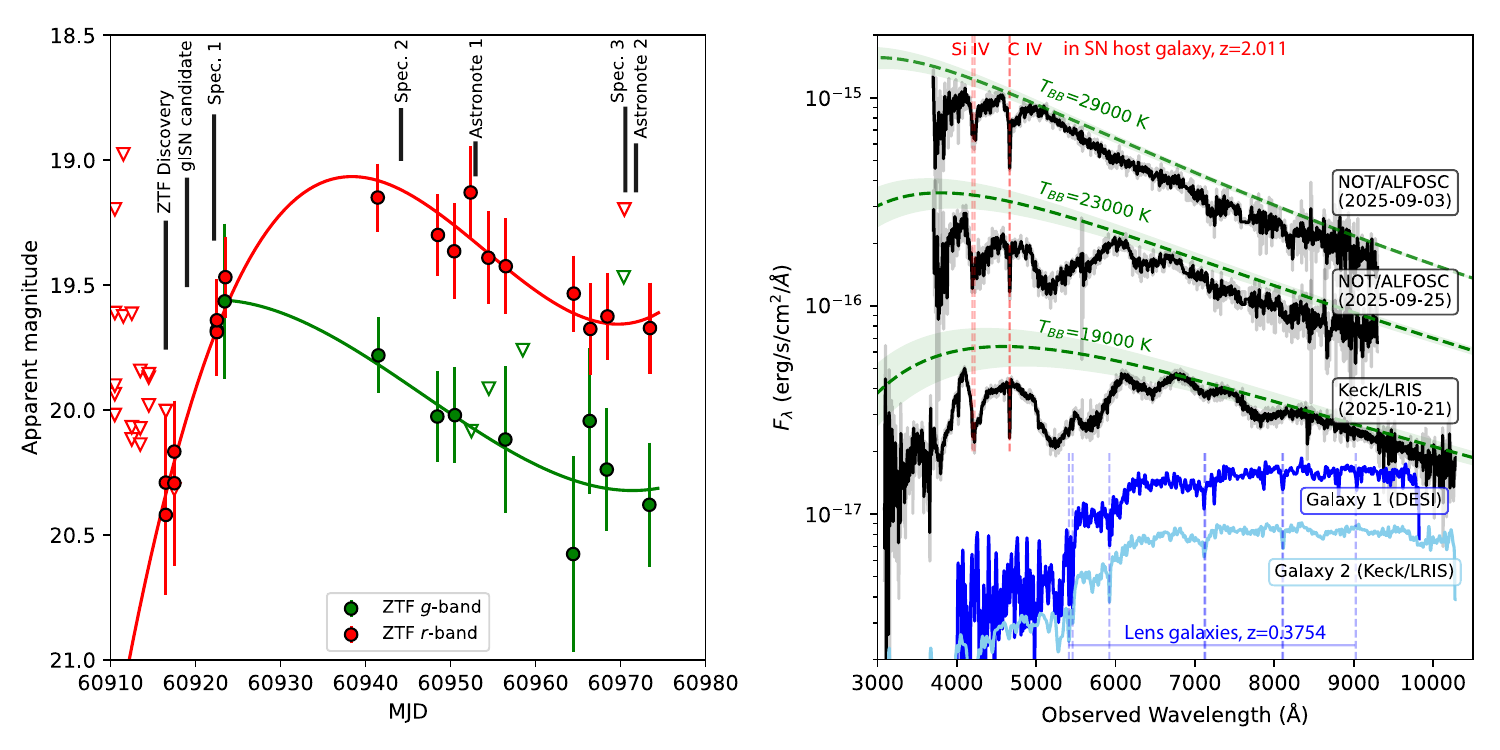}
    \caption{ {\bf Left panel}: Public ZTF $g$ and $r-$band light curves of SN\,2025wny, shown as green and red symbols. Black vertical lines indicate the time of the discovery, spectroscopic observations and our reported AstroNotes \citep{wise2025,johansson2025}.
    {\bf Right panel}: Spectra of SN\,2025wny (black lines) and the two galaxies forming the lens system (blue lines). Prominent  absorption lines from the SN host galaxy at $z=2.011 \pm 0.001$, and the two lens galaxies at $z_{\rm lens}=0.3754$ are marked with vertical red and blue lines, respectively. The dashed green curves show blackbody fits to the SN spectra, with $T_{BB} \sim$29000, 23000 and 18000 K, for the three spectra. 
    }
    \label{fig:all_spec}
\end{figure*}

\subsection{Spectroscopy \label{sec:obs_spec}}

Three spectra of SN\,2025wny are presented in this work\footnote{Additional spectra were acquired with the the Next Generation Palomar Spectrograph (NGPS) on the Palomar 5m Hale telescope, with the Kait Spectrograph on the Lick 3m Shane telescope, and with the Spectral Energy Distribution Machine (SEDM) on P60; these spectra will be presented in future work.}: two obtained using the Alhambra Faint Object Spectrograph and Camera (ALFOSC) on the 2.56m Nordic Optical Telescope (NOT) on UT 2025-09-04 and 2025-09-26, and one with the Low-Resolution Imaging Spectrometer (LRIS, \citealt{Oke95}) on the Keck I telescope on UT 2025-10-22. 

The NOT/ALFOSC spectra were obtained with a 1\farcs0 wide slit using Grism 4. We reduced the spectra with a custom fork of \texttt{PypeIt} \citep{pypeit:zenodo,pypeit:joss_arXiv,pypeit:joss_pub}, following standard procedures for preprocessing, 1D extraction, wavelength calibration, and flux calibration. 

The Keck/LRIS observations were obtained with a 1\farcs0 wide slit using the B400/3400 grism and R400/8500 grating at a position angle (PA) of $164^{\circ}$ (covering Images A and B, $45^{\circ}$ away from the parallactic angle) as well as additional spectra at other slit orientations covering other SN images and ``Galaxy B''.  Two exposures of 600\,s were taken at each slit position.  Spectra were reduced and extracted using \texttt{LPipe} \citep{2019PASP..131h4503P}. 

The first NOT spectrum was almost entirely featureless except for a narrow absorption line at 4664\,\AA\ and a possibly broader absorption feature at 4178~\AA, not matching any known lines that were considered likely at the time.  By the time of the second spectrum two weeks later, clear broad SN-like features had developed in the blue half of the spectrum, although these did not match any optical SN spectral templates.

The LRIS observation revealed numerous additional narrow absorption lines matching strong intergalactic/interstellar features at a common redshift of $z=2.011 \pm 0.001$, establishing the redshift of the event. (The narrow line at 4664\,\AA\ previously seen in the NOT spectra was one such feature, the \ion{C}{4} $\lambda\lambda$1550 doublet). 
Comparing the spectrum to UV templates, the broader features in the spectrum were found to match Type I superluminous SNe, as reported by \citet{johansson2025}.  More details on the spectral analysis are provided in \S \ref{sec:spec_analysis}.

\section{Spectral Analysis \label{sec:spec_analysis}}

\begin{figure}
    \centering
    \includegraphics[width=\columnwidth]{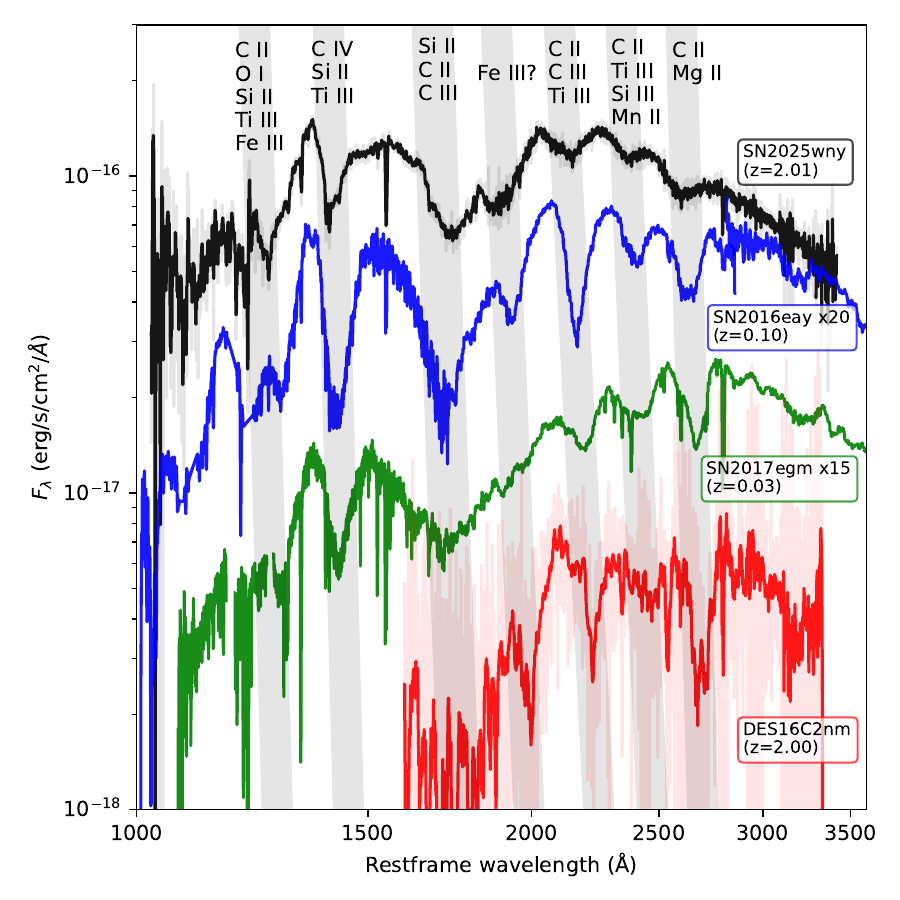}
    \caption{Keck/LRIS spectrum of SN\,2025wny (black line) compared to HST spectra of the low-$z$ SLSN-I 2016eay (a.k.a Gaia16apd, blue line), SN\,2017egm (green line) and a Magellan/LDSS3 spectrum of high-$z$ DES16C2nm (red line). All spectra are scaled by their luminosity distance relative to SN\,2025wny, and for clarity SNe 2016eay and 2017egm have been offset by multiplicative factors of 20 and 15, respectively.
    }
    \label{fig:slsn_comparison}
\end{figure}

A comparison between the LRIS spectrum of SN\,2025wny and rest-frame UV spectra of three comparison superluminous supernovae is shown in Figure \ref{fig:slsn_comparison}. The comparison events are SN\,2016eay and SN\,2017egm (two nearby Type I SLSNe observed with $HST$; \citealt{Yan2017ApJ...840...57Y,Kangas2017MNRAS.469.1246K,Yan2018ApJ...858...91Y}), and DES16C2nm (a Type I SLSN at $z=1.998$ that is not strongly lensed; \citealt{Smith2018ApJ...854...37S}).  All four spectra were taken shortly after the rest-frame near-UV peak of the light curve (\S \ref{sec:phot_analysis}).  
The locations of the broad absorption troughs seen in SN\,2025wny match similar features in all three comparison objects, and can be attributed to various metals (primarily C, Mg, Si, Ti, Fe), as in previous studies of SLSNe-I \citep{quimby_spectra_2018}. The spectra were not good matches to other potential UV-luminous comparison objects (e.g., TDEs, SLSNe-II, or QSOs). This, combined with the blue colors, establishes SN\,2025wny as a Type I superluminous supernova (SLSN-I). 

While the spectrum of SN\,2025wny resembles that of comparison SLSNe-I, it exhibits a bluer continuum than any of the comparison objects and the absorption features are generally weaker, with broader profiles. 
Fitted blackbody temperatures (using the near-UV only to avoid line blanketing) range from 29000$\pm$1000 to 19000$\pm$1000 K, somewhat higher than what has been inferred from spectral fitting of other SLSNe with UV spectra \citep{Yan2017ApJ...840...57Y}.

\begin{figure}
    \centering
    \includegraphics[width=\columnwidth]{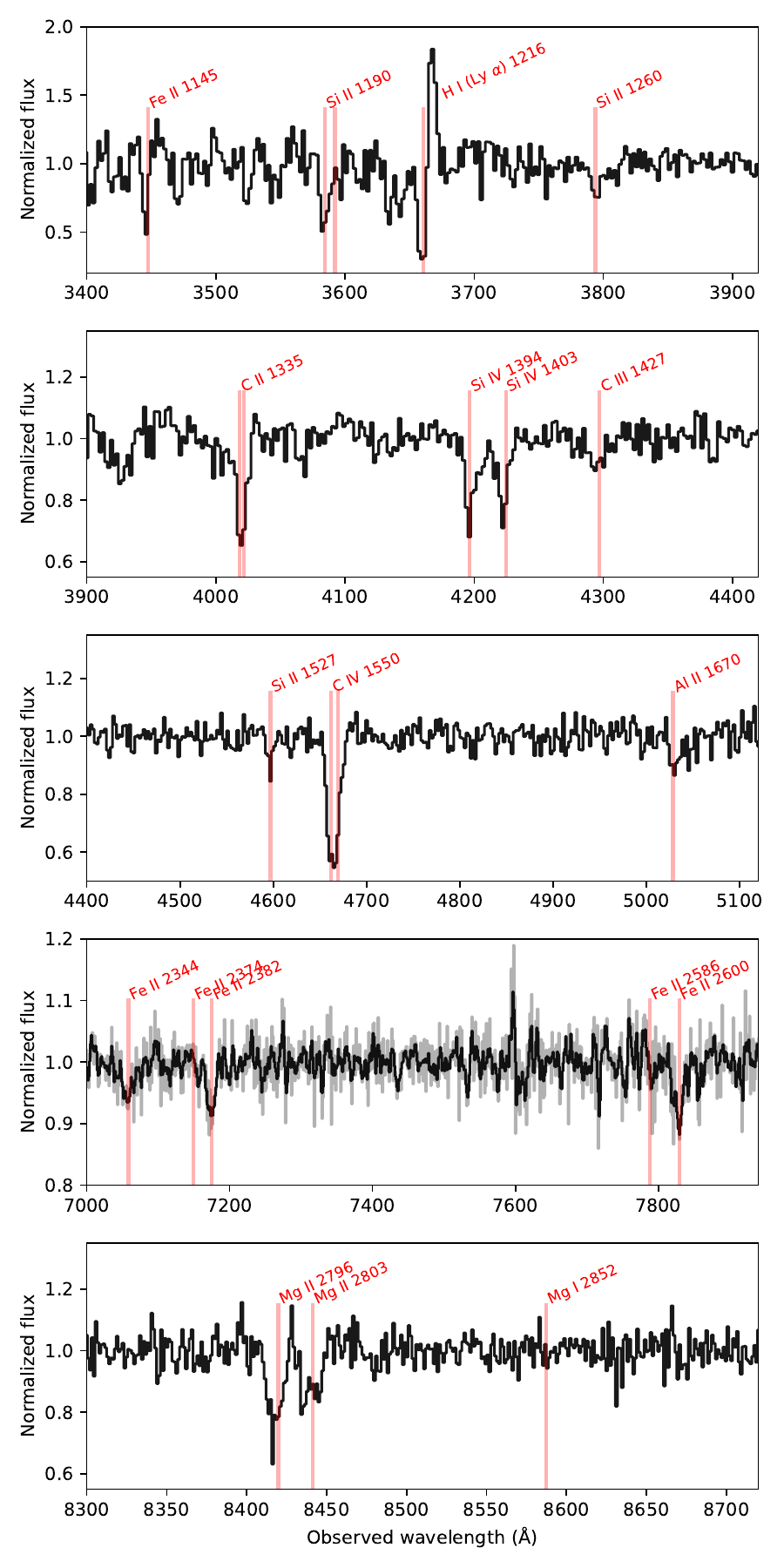}
    \caption{Selection of absorption lines identified for SN\,2025wny, used to determine the SN host redshift $z=2.011 \pm 0.001$.}
    \label{fig:abs_features}
\end{figure}

The spectrum also shows a rich series of narrow absorption lines associated with absorption from the host galaxy. \autoref{fig:abs_features} shows the spectrum after division by a continuum normalization model, with key detected features including 
\ion{H}{1} $\lambda$1216 (Ly$\alpha$), \ion{Si}{2} $\lambda$1260, \ion{C}{2} $\lambda$1335, \ion{Si}{4} $\lambda\lambda$1394,1403, \ion{Si}{2} $\lambda$1527, \ion{C}{4} $\lambda$1550, and \ion{Mg}{2} $\lambda$2800. No significant absorption features from the $z=0.3754$ lensing system are evident (including \ion{Fe}{2} 2600 and or \ion{Mg}{2} 2800, which are within our spectral coverage but not seen).

While a detailed study of the narrow absorption lines is beyond the scope of this analysis, we note that their strengths are relatively weak compared to what has been seen in other high-redshift absorption systems.  While Mg II is detected with equivalent width comparable to what has been seen in other SLSNe, most other lines are weaker and the \ion{Fe}{2}\,$\lambda$\,2344, 2383, 2600 series commonly seen in GRB afterglow spectra \citep{Selsing2019a} is only marginally detected.  We extracted equivalent widths of the features marked on Figure \ref{fig:abs_features} and used these to compute the mean line-strength parameter (LSP) of \citet{deUgartePostigo2012a}.  We measure LSP~=~$-0.64$, placing the absorption strength at the 20th percentile of the GRB distribution.  
These suggests a low-density sightline and/or a low galaxy metallicity.

A particularly notable case is the Lyman-$\alpha$ line: this line shows no damped wings and is consistent with being unresolved at the resolution of the spectrograph.  This implies that the neutral hydrogen column is quite low: a conservative upper limit from fitting a Voigt profile is log$_{10}$($N_{H}$/cm$^{-2}$) $<$ 19.3, which is already less than all but a few of the lowest-$N_H$ GRB sightlines \citep{Tanvir+2019} and essentially all low-$z$ star-forming galaxies for which a Lyman-$\alpha$ absorption measurement has been possible \citep{McKinney+2019,Kulkarni+2022}. A narrow Lyman-$\alpha$ emission component is also detected in the spectrum of SN\,2025wny just redward of the absorption feature, likely originating from star-formation in the host.  These properties are indicative of a low-mass star-forming galaxy with a low neutral gas density and little dust extinction.

\begin{figure}
    \centering
    \includegraphics[width=\columnwidth]{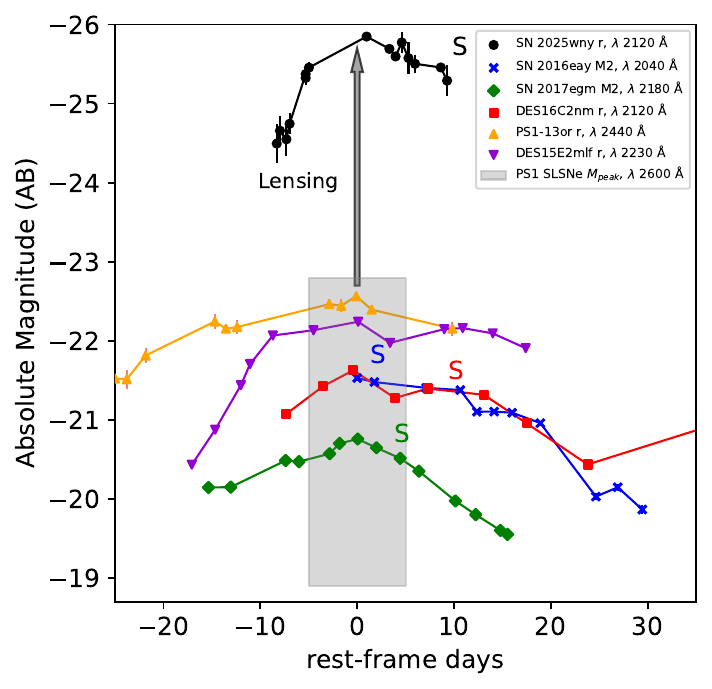}
    \caption{Rest-frame UV light curve of Image A of SN\,2025wny (from the ZTF $r$-band) compared to a set of comparison SLSN curves in similar rest-frame bands: SN\,2017egm, SN\,2016eay, PS1-13or, DES15E2mlf, and DES16C2nm.  The observation times of the spectra used for the comparison in Figure \ref{fig:slsn_comparison} are marked with an "S".
    } 
    \label{fig:lightcurve}
\end{figure}

\section{Light Curve Analysis and Magnification \label{sec:phot_analysis}}

The $r-$band light curve from ZTF is presented again in Figure \ref{fig:lightcurve}. The vertical axis shows the rest-frame absolute magnitude at the equivalent rest-frame central wavelength ($\lambda = 2120$\,\AA) with correction for Galactic extinction \citep{Schlafly+2011} but no correction for host extinction or lensing magnification. 
For comparison, the host-subtracted Swift Ultraviolet-Optical Telescope (UVOT; \citealt{Roming+2005}) light curves\footnote{UVOT light curves were built directly using data from the \emph{Swift} archive using HEAsoft 6.30.} of two low-$z$ SLSNe (SN\,2016eay at $z=0.10$ and SN\,2017egm at $z=0.03$) are also shown in the approximately matching UVM2 filter.  Also shown are a few examples of very luminous SNe found at higher redshift in deeper surveyssimilarly matched by rest-frame central wavelength: PS1-13or from \cite{lunnan_hydrogen-poor_2018}, DES15E2mlf from \cite{Pan2017MNRAS.470.4241P}, and DES16C2nm from \cite{Smith2018ApJ...854...37S}.  The gray box indicates the range of peak luminosities standardized at a (slightly redder) wavelength of $\lambda=2600$\,\AA\ from \cite{lunnan_hydrogen-poor_2018}.

Due to time dilation effects, the two-month period between the discovery of SN\,2025wny and the most recent data point corresponds to only 20 days in the rest frame, so only a limited timespan is available to characterize its evolution.  The rise of 1.5 mag over the first 10 rest-frame days is relatively steep for a SLSN (although not unprecedented: the light curve of DES15E2mlf shows similar behavior). 

At a redshift of $z=2.011$, the peak absolute magnitude of the SN without lensing correction is approximately $-25.8$ at $\lambda_{\rm rest} = 2120$\,\AA. If SN\,2025wny is intrinsically of similar luminosity and host extinction to a ``typical'' low-$z$ SLSN (such as SN\,2016eay), it would require a lensing magnification of $\sim$4.3 mag (factor $\mu \sim 50$). Comparing instead to PS1-13or, the most UV-lumninous event in the Pan-STARRS sample of \cite{2018ApJ...852...81L}, 
would still require a magnification of $\sim$3.2 mag ($\mu \sim 20$).

We infer that at least one, and possibly all, of the following are true: \emph{(a)} the magnification factor of the system is very large ($\mu \gtrsim 20$, and possibly as much as $\mu \sim 50$);
\emph{(b)} in addition to being lensed, SN\,2025wny was a particularly overluminous SLSN in the ultraviolet, or \emph{(c)} SN\,2025wny was discovered and observed in the rest-frame UV at a much earlier phase than is normally possible for these events, and the early temperatures and UV luminosities for some SLSNe are much higher than previously anticipated.  Future observations and analysis will help to distinguish these possibilities.

\section{Conclusions}

The discovery of SN\,2025wny underscores the accelerating pace at which lensed supernovae are being found, marking the transition from rare, serendipitous detections to an emerging population accessible through systematic time-domain surveys. Over the next decade, this field is poised for rapid growth. The Vera C. Rubin Observatory's Legacy Survey of Space and Time (LSST) will transform the landscape of lensed transient discovery, with forecasts predicting several tens of multiply imaged SNe per year across a broad range of redshifts and environments \citep[e.g.,][]{goldstein_2019, wojtak_2019, arendse_2024, 2024MNRAS.535.2523S}. LSST’s depth, cadence, and photometric stability will enable the detection of faint, highly magnified events, while its extensive sky coverage will build the statistical foundation necessary to test models of galaxy mass distributions and cosmological parameters.

At the same time, space-based missions such as JWST and Euclid will provide complementary imaging and spectroscopy crucial for time-delay measurements, host-galaxy characterization, and high-resolution lens modeling. In particular, JWST/NIRSpec and NIRCam observations can yield precise phase-resolved spectra and astrometry of individual SN images, such observations, along with HST imaging in the optical are scheduled for SN\,2025wny (PI: Goobar). In the longer term, Euclid’s wide-area lens catalog will greatly expand the pool of candidate deflectors \citep{2015ApJ...811...20C}. Together with ongoing efforts by ground-based facilities, this synergy will enable routine discovery and detailed study of lensed SNe, bridging the gap between current pilot samples and the statistically powerful datasets required for cosmology.

As analysis techniques continue to mature---incorporating forward modeling, simulation-based inference, and improved microlensing treatments---lensed SNe are expected to evolve from curiosities to precision probes of both lens physics and cosmic expansion. The discovery of SN\,2025wny thus represents not only an individual milestone but also a glimpse into a rapidly unfolding era of time-delay cosmography powered by the next generation of time-domain surveys.

In addition to underscoring the utility of lensed superluminous supernovae as cosmological probes, the discovery of SN\,2025wny also opens up a new pathway for the use of these events to study galaxy evolution.  While the potential for very luminous supernovae to serve as probes of high-$z$ dwarf galaxies was realized shortly after the discovery of the first few events at $z>1$ \citep{Berger2012a}, in practice SLSNe have rarely been able to serve this purpose due to the large amounts of telescope time required to obtain meaningful S/N beyond $z>1$.  While our observations so far do not permit a detailed analysis of the host environment yet, deeper observations with higher-resolution instruments would permit constraints on the metallicity and kinematics of the host ISM that could then be complemented by detailed late-time observations of the (also magnified) host itself, providing an unprecedented view into the nature of a distant star-forming galaxy in both absorption and emission.

\vspace{1.5cm}

We thank Stefan Taubenberger for sharing the HOLISMOKES team's pre-submission draft on this object \citep{2025arXiv251021694T}, following our AstroNote on the classification of SN\,2025wny.

Based on observations obtained with the Samuel Oschin Telescope 48-inch and the 60-inch Telescope at the Palomar Observatory as part of the Zwicky Transient Facility project. ZTF is supported by the National Science Foundation under Award 2407588 and a partnership including Caltech, USA; Caltech/IPAC, USA; University of Maryland, USA; University of California, Berkeley, USA; University of Wisconsin at Milwaukee, USA; Cornell University, USA; Drexel University, USA; University of North Carolina at Chapel Hill, USA; Institute of Science and Technology, Austria; National Central University, Taiwan, and OKC, University of Stockholm, Sweden. Operations are conducted by Caltech's Optical Observatory (COO), Caltech/IPAC, and the University of Washington at Seattle, USA. 

The Liverpool Telescope is operated on the island of La Palma by Liverpool John Moores University in the Spanish Observatorio del Roque de los Muchachos of the Instituto de Astrofisica de Canarias with financial support from the UK Science and Technology Facilities Council.

Based on observations made with the Nordic Optical Telescope, owned in collaboration by the University of Turku and Aarhus University, and operated jointly by Aarhus University, the University of Turku and the University of Oslo, representing Denmark, Finland and Norway, the University of Iceland and Stockholm University at the Observatorio del Roque de los Muchachos, La Palma, Spain, of the Instituto de Astrofisica de Canarias. The NOT data were obtained under program ID P70-501.

Some of the data presented herein were obtained at the W.~M. Keck Observatory, which is operated as a scientific partnership among the California Institute of Technology, the University of California, and NASA. The Observatory was made possible by the generous financial support of the W.~M. Keck Foundation. The authors wish to recognize and acknowledge the very significant cultural role and reverence that the summit of Maunakea has always had within the indigenous Hawaiian community. We are most fortunate to have the opportunity to conduct observations from this mountain.



\facilities{Keck:I (LRIS), NOT (ALFOSC), Liverpool:2m, PO:1.2m}

\software{Astropy \citep{astropycollaboration_astropy:_2013, astropycollaboration_astropy_2018, AstropyCollaboration2022}, 
Matplotlib \citep{hunter_matplotlib:_2007}, 
}
\normalsize
\bibliography{references, Lensbib}

\end{document}